\documentclass{iaus}
\usepackage{graphicx}
\def\sna{SN\,Ia}
\def\apj{ApJ}
\def\apjl{ApJ Lett.}
\def\apjs{ApJS}

\def\apss{Ap\&SS}
\def\mnras{MNRAS}
\def\nat{Nature}
\def\aap{Astron. Astrophys.}
\def\apgt{\ {\raise-.5ex\hbox{$\buildrel>\over\sim$}}\ }
\newcommand{\ms}{\mbox {$M_{\odot}$}}
\newcommand{\mch}{\mbox {$M_{Ch}$}}
\newcommand{\myr}{\mbox {~$\rm M_{\odot}$~yr$^{-1}$}}
\newcommand{\porb}{\mbox {$P_{\rm orb}$}}
\newcommand{\md}{\mbox {$\dot{M}$}}

\title[SN\,Ia and Supersoft X-ray Sources ] 
{Type Ia Supernovae and Supersoft X-ray Sources }

\author[L.R. Yungelson]   
{L.R. Yungelson}

\affiliation{Institute of Astronomy, RAS,\\
48 Pyatnitskaja Str., 119017 Moscow, Russia \\ email: {\tt lev.yungelson@gmail.com} }

\pubyear{2011}
\volume{281}  
\pagerange{1---4}
\date{August 23, 2011}
\jname{Binary Paths to Type Ia Supernovae Explosions }
\editors{R. Di Stefano \& M. Orio eds.}

\begin{document}

\maketitle

\begin{abstract}
The rates of \sna\ 
for double-degenerate and 
single-degenerate  
scenario are computed for the 
models of spiral and  
elliptical galaxies. The number of nuclearly burning white dwarfs (NBWD) is traced. 
The data favours double-degenerate
scenario 
and suggests lower number  
of  NBWD per unit mass in ellipticals and their lower average mass 
as one of the reasons for the difference in the number 
of supersoft X-ray sources observed in the galaxies of different types.
\end{abstract}

\firstsection 
\section{Introduction}
While there exists  an agreement
that \sna\ result from thermonuclear explosions 
of  white dwarfs (WD) which accumulated Chandrasekhar mass $\mch\approx1.38\ms$\ 
(\cite[Hoyle \& Fowler 1960]{1960ApJ...132..565H}), the process of accumulation  
of \mch\ remains an enigma. 
It was suggested that WDs
may accumulate \mch\ via accretion in semidetached binaries 
(single-degenerate scenario, SD, \cite[Schatzman 1963]{1963stev.conf..389S}), 
merger of binary WDs with $M_1+M_2\apgt\mch$\ (double-degenerate scenario, DD, 
\cite[Webbink 1979]{1979wdvd.coll..426W}),
accretion of wind matter in symbiotic stars 
(\cite[Truran \& Cameron 1971]{1971Ap&SS..14..179T}).
It is also hypothesized that accretion of  He
onto sub-Chandrasekhar CO WD  may lead to \sna\  via double-detonation  
\cite[(Livne 1990)]{1990ApJ...354L..53L}.

However, at the moment, only SD- and DD-scenario may be discussed
as viable ones. Accretion in wide symbiotic systems is most probably 
not efficient enough for accumulation of \mch. 
Double-detonation 
scenario, though claimed to provide ``correct'' delay-time distribution of \sna\ 
(DTD) and enable a significant fraction of the total \sna\
rate \cite[(Ruiter et al. 2011)]{2011MNRAS.tmp.1282R},  
very strongly depends on the
assumed efficiency of He-accumulation (Piersanti et al., these proceedings).
Also, models do not fit observables, due to the presence of He-burning 
products in the outer layers of ejecta.  

The count 
of supersoft X-ray sources (SSS) 
in external galaxies may provide certain clues to the problem of progenitors of \sna\
\cite[(Di Stefano 2010,]{2010ApJ...712..728D} 
\cite[Gilfanov \& Bogd{\'a}n 2010)]{2010Natur.463..924G}.     
SSS, which are deemed to be nuclearly burning white dwarfs (NBWD, 
\cite[van den Heuvel et al. 1992)]{1992A&A...262...97V},
may be direct precursors of \sna\ in 
the SD-scenario. In the DD-scenario,
a binary may be a SSS
when the first-formed WD accretes matter from the stellar wind of 
the precursor of the second WD.         
The number of observed SSS provides the lower limit of the number of  
NBWD, since the latter do not 
necessarily radiate in X-ray.

Below, we consider the relations between the rates of formation of NBWD, \sna, SSS in
the model of stellar system with a constant for 10\,Gyr star
formation rate  $\dot{M}_{\star}=8\,\myr$ and 
for a model in which the same mass of stars is formed 
in a 1\,Gyr long star-formation burst, i.e. for toy-models of a spiral and an elliptical 
galaxy. 
We test also the influence of CE parameters and mass accumulation 
efficiency upon DTD.
Our assumptions are described  in  \cite[Yungelson (2010)]{2010AstL...36..780Y}. 
Note, we do not consider effects like ``stripping'' 
or ``companion reinforced attrition process''.

\begin{figure}[t]
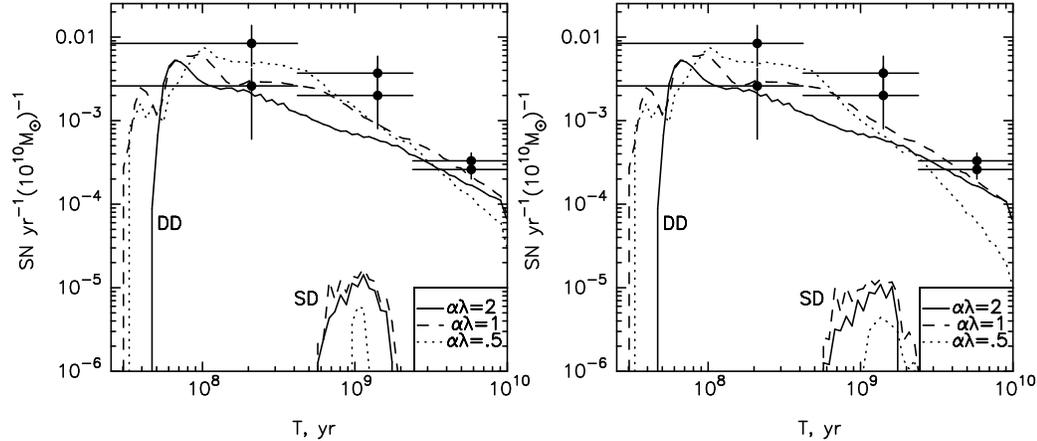

\begin{center}
\begin{minipage}[t]{0.49\textwidth}
 \includegraphics[width=2.3in,angle=-90]{fig1a.eps}
\end{minipage}
\begin{minipage}[t]{0.49\textwidth}
 \includegraphics[width=2.3in,angle=-90]{fig1b.eps}
\end{minipage}
 \caption{Model DTD for DD- and  SD-scenarios compared to the
 observational data  \cite[(Maoz et al. 2011)]{2011MNRAS.412.1508M}. Two sets of 
the latter correspond to different models of the dust. 
Left panel --- H- and He-accumulation  efficiency after 
\cite[Prialnik \& Kovetz (1995)]{1995ApJ...445..789P} and
\cite[Iben \& Tutukov (1996)]{1996ApJS..105..145I},
right panel --- efficiency after \cite[Hachisu et al. (1999)]{1999ApJ...522..487H}
and \cite[Kato \& Hachisu (2004)]{2004ApJ...613L.129K}.
}  
   \label{fig:dtd}
\end{center}
\end{figure}
\begin{figure}[t]
\begin{center}
 \includegraphics[scale=0.8,angle=-90]{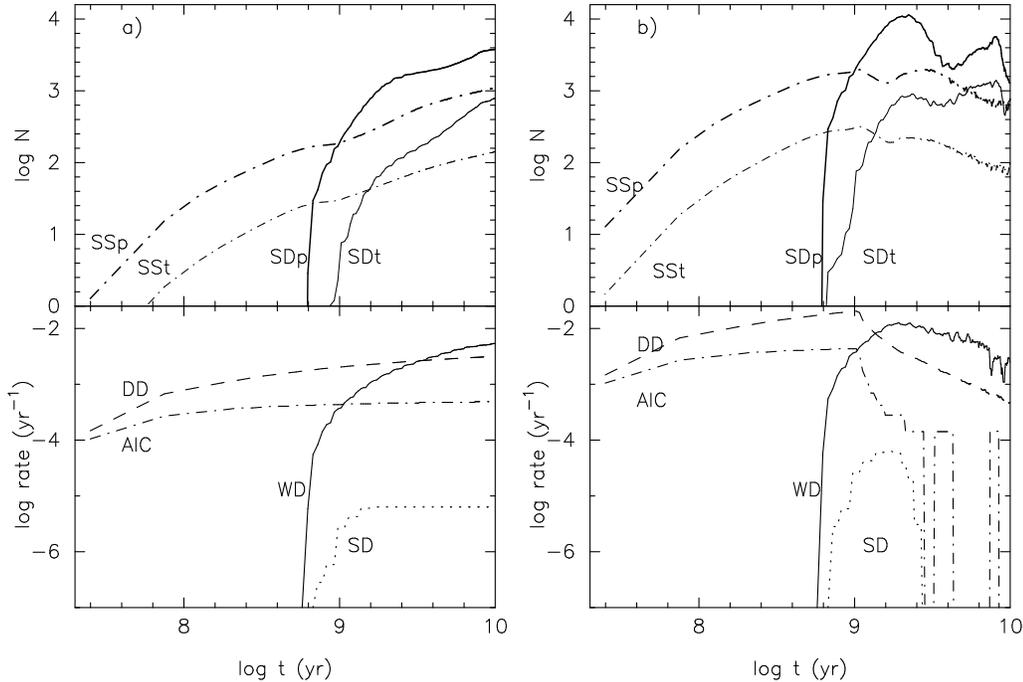}
  \caption{Panel (a), lower part --- evolution of the rates of \sna\ in DD-scenario 
(dashed line) and in 
SD-scenario (dotted line), formation of NBWD (solid line), AICs in symbiotic stars 
in the model of a ``spiral'' galaxy. Upper part --- evolution of the numbers of 
permanent (SSp) and transient (SSt) SSS in symbiotic stars and permanent (SDp) and 
transient (SDt) sources in semidetached systems. Panel (b) shows evolution of 
the same rates and numbers in the model of an ``elliptical'' galaxy.}  
   \label{fig:rates}
\end{center}
\end{figure}

\section{Results}
In Fig. \ref{fig:dtd} we compare DTD for DD- and SD-scenarios obtained 
for different
values of the common envelope (CE)  efficiency and donor envelope binding energy 
parameter product
$\alpha_{ce}\times\lambda$  in \cite[Webbink's (1984)]{1984ApJ...277..355W} 
equation for CE. Total dominance of DD mergers is clear,
irrespective of assumptions on  $\alpha_{ce}\times\lambda$ and efficiency of matter 
accumulation. The main reason for this is a narrow range of combinations of 
masses of components in semidetached systems 
which allows steady accumulation of the matter. Bearing in
mind all uncertainties involved in population synthesis and 
in derivation of DTD from observations, 
the slope of the model DD
DTD ($t^{-0.8}$) compares well with the slope of DTD curve ($t^{-1.2\pm0.3}$ at 
$t \apgt 400$\,Myr) derived by \cite[Maoz et al. (2011)]{2011MNRAS.412.1508M} 
for \sna\ at $0<z<1.45$.

Figure \ref{fig:rates} shows
results of computations for two model stellar systems.   
Formation rate of close binary WD reflects SFR, hence, \sna\ in the DD-scenario
start at the age of several tens of Myr
when the most massive pairs of WD begin to form. 
In the  ``spiral'' galaxy binary WD form continuously and DD-\sna\
rate permanently increases, since merge both 
``old'' initially relatively wide pairs and ``young'' relatively close ones. 
In the ``elliptical'' galaxy the reservoir of binary WD 
created in the star-formation burst gradually ``melts'' 
and soon after cessation of star formation DD-\sna\ rate starts to decline.

In semidetached systems \sna\ are delayed by  
 $\sim\,10^{9}$\,yr respective to star formation. 
After a star-formation burst, semidetached systems in which 
a CO WD is able to accumulate 
\mch\ form and exist only over a limited  time span of several Gyr
\cite[(Canal et al. 1996 and numerous later papers)]{1996ApJ...456L.101C}. 
In our model, \sna\ occur in binaries with 
 $M_{wd,0} \apgt 0.85$\,\ms\ and  
$M_{sg,0} \apgt 1.4$\,\ms\ at the beginning of Roche-lobe overflow.  
Figure~\ref{fig:rates} shows 
that formation rate of semidetached systems with NBWD able to accumulate \mch\ 
is only a minor fraction of the total 
formation rate of systems with NBWD 
even in the relatively early stages of evolution. 
Semidetached systems with NBWD form continuously, but 
\textit{in the case of burst-like star formation currently observed SSS are not 
precursors of} \sna. The epoch of \sna\ in semidetached systems is short:
in the model ``spiral'' galaxy at $t=10$\,Gyr \sna\ 
occur in stars formed approximately  
600\,Myr to 2.5\,Gyr ago.

The phenomenon of symbiotic stars is associated predominantly with wide binaries, 
in which WD form neither through RLOF nor CE. 
As the donor begins to ascend the giant branch, its stellar wind is weak and in a 
binary with $\porb\sim(100-1000)$\ day 
accretion is extremely inefficient.
Newborn NBWD first becomes an unstable burner and, typically, it
may become a stationary burner only shortly 
before the loss of the envelope by companion  
\cite[(Yungelson et al. 1995)]{1995ApJ...447..656Y}.
The amount of matter which may be retained by WD is as a rule 
$\lesssim 0.1$\,\ms\   
\cite[(L{\"u} et al. 2006)]{2006MNRAS.372.1389L}. An implication of this is that CO WD 
with $M_0\lesssim 1.1-1.2$\,\ms\ 
hardly reach \mch, while more massive ONeMg WD experience 
accretion-induced collapses (AIC, \cite[Nomoto \& Kondo 1991]{1991ApJ...367L..19N}). 
AIC's may explain the weakest peculiar \sna\ 
\cite[(Metzger et al. 2009)]{2009MNRAS.396.1659M}. Evolution of the rate of AIC's in the 
models is also shown in Fig.~\ref{fig:rates}.

\begin{figure}[t]
\begin{center}
 \includegraphics[scale=0.4,angle=-90]{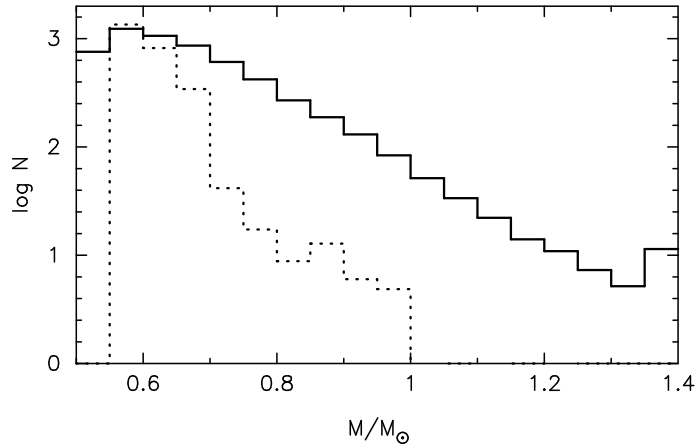}
  \caption{Mass spectrum of NBWD in the model systems with continuous star formation 
(solid line) and with initial starburst  (dotted line). 
The histograms show  total contribution of semidetached and symbiotic 
systems. }  
   \label{fig:masses}
\end{center}
\end{figure}

In Fig.~\ref{fig:masses} we show the distribution of    
NBWD over masses in two models. Distributions sum up both transient and permanent 
burners. At any time the latter dominate due to longer lifetimes.    
In the systems with subgiants the fraction of outbursting systems is initially low, 
since first start to overfill their Roche lobes relatively massive stars with high 
\md. Later, the fraction of systems in which the rate of accretion is lower than 
the limit 
for stable H-burning increases, since the systems with massive donors finish their 
evolution fast.
At $t$=10\,Gyr, the number of NBWD in a ``spiral'' galaxy exceeds their number 
in an ``elliptical'' galaxy of the same mass and masses of WD are also higher 
(Fig. \ref{fig:masses}).
Relatively well studied nearby spiral galaxies 
contain $\sim$ 100 SSS each, 
while elliptical galaxies 
host only $\sim10$  SSS per galaxy \cite[(Di Stefano 2010)]{2010ApJ...712..728D}.  
Taking into account that \textit{Chandra}
is able to observe in these spiral galaxies only SSS with  
$M_{wd}\apgt (1.0 - 1.2)$\,\ms\ and 
objects with $M_{wd} \apgt 0.8$\,\ms\ in elliptical galaxies 
\cite[(Di Stefano 2010)]{2010ApJ...712..728D}, the number of NBWD in the models
at least qualitatively
agrees with the one expected from observations. 
This brings to conclusion that the model in which the rate of \sna\ 
is defined by DD-mechanism and which is reasonably 
consistent with observationally inferred DTD, is
also able to explain,  at least partially, the difference in the numbers of SSS 
in spiral and elliptical galaxies.

\medskip
The author acknowledges  support from IAU, 
RFBR (grant No. 10-02-00231), 
and the Program of the Russian Academy of Sciences 
``Origin and evolution of stars and galaxies''.

\end{document}